\documentclass[conference]{IEEEtran}
\IEEEoverridecommandlockouts
\usepackage{caption}
\usepackage{graphicx}
\usepackage{subfigure}
\usepackage{lipsum}
\usepackage{url}
\usepackage{verbatim}
\usepackage{color}

\usepackage{graphics}
\usepackage{amsmath,amsfonts, amssymb}
\usepackage{mathtools}
\usepackage{epstopdf}
\usepackage{amsthm}
\usepackage{bbm, bm}
\usepackage{algorithm}
\usepackage[noend]{algpseudocode}
\usepackage{dsfont}
\usepackage{cite}
\usepackage{multicol}
\usepackage{multirow}
\usepackage{array}
\usepackage{booktabs}
\usepackage[normalem]{ulem}
\useunder{\uline}{\ul}{}

\newcolumntype{P}[1]{>{\centering\arraybackslash}p{#1}}
\newcolumntype{M}[1]{>{\centering\arraybackslash}m{#1}}

\def\BibTeX{{\rm B\kern-.05em{\sc i\kern-.025em b}\kern-.08em
    T\kern-.1667em\lower.7ex\hbox{E}\kern-.125emX}}

\begin{document}

\title{Do not Interfere but Cooperate:\\ A Fully Learnable Code Design for \\ Multi-Access Channels with Feedback \\
}

\author{\IEEEauthorblockN{Emre Ozfatura, Chenghong Bian and  Deniz G{\"u}nd{\"u}z}
\IEEEauthorblockA{~Information Processing and Communications Lab (IPC-Lab), Imperial College London, London, UK\\
    Email:$\{$m.ozfatura, c.bian22, d.gunduz$\}$@imperial.ac.uk}}
\maketitle

\begin{abstract}
Data-driven deep learning based code designs, including low-complexity neural decoders for existing codes, or end-to-end trainable auto-encoders  have exhibited impressive results, particularly in scenarios for which we do not have high-performing structured code designs.  However, the vast majority of existing data-driven solutions for channel coding focus on a point-to-point scenario. In this work, we consider a multiple access channel (MAC) with feedback and try to understand whether deep learning-based designs are capable of enabling coordination and cooperation among the encoders as well as allowing error correction. Simulation results show that the proposed multi-access block attention feedback (MBAF) code improves the upper bound of the achievable rate of MAC without feedback
in finite block length regime.

\end{abstract}


\section{Introduction}

The design of deep learning (DL) based solutions for wireless communication problems has become a rapidly evolving research area. It has been shown that data-driven solutions are either able to outperform their hand-crafted counterparts, or exhibit comparable performance with lower computational complexity. In the context of error correction coding, it has been shown  by various recent studies \cite{TurboAE,turbo_int,deepcode,ND_res,kocode,ListAE,NBP_LBCD} that DL-assisted designs can be trained and employed to achieve reliable communication as an alternative approach to well-known structured codes. These designs can be grouped into two main categories based on the stage/phase that data-driven design is incorporated. The first category includes solutions where the decoding of conventional codes, e.g., LDPC codes and polar codes, is done using deep neural network (DNN) architectures, e.g., neural belief propagation \cite{ND_res,cyclic,graph_decode}. The works in the second category, on the other hand, employ DNN-based modules both at the encoding and decoding sides by introducing a fully learnable auto-encoder (AE) architecture  \cite{TurboAE,ECCT,BCWF,kocode}. The experimental results of aforementioned works highlight that, DNN-based decoder/AE models either achieve the state-of-the-art block error rate (BLER) performance in the finite block length regime, or achieve similar BLER performance with conventional encoding/decoding schemes with lower complexity.\\
\indent The use of DNNs  is not limited to  the feed-forward error correction problem; recently, it has been shown in \cite{gbaf, deepcode, AttentionCode, baaf, deepcode_act} that DNN-based solutions have significant potential when a feedback link from the receiver to the transmitter is available. Indeed, the feedback channel, highly relevant for practice, is considered as a challenging model and practical code designs have been mostly limited to
repetition based, e.g., ARQ/HARQ, protocols. On the other hand, experimental results in \cite{gbaf, baaf} demonstrate that DNN-based feedback code designs can provide significant performance gains in the short block length regime compared to hand-crafted solutions.\\
\indent Although the advantages of  DNN-based solutions have been shown in point-to-point systems with/without feedback scenarios, we have a limited understanding of their role in multi-user scenarios. The broadcast channel scenario has been recently analyzed in \cite{BCWF}. In this paper, we focus on the multiple access channel (MAC) with feedback, and analyze the applicability of DNN-based solutions. It is known that the existence of channel feedback enlarges the capacity region of the MAC; however,  the design of practical codes, especially for the short block length regime, is still an open problem. To the best of our knowledge, this is the very first work that studies a fully-learnable framework for the MAC with feedback scenario. We remark that, in the classical MAC, since the transmitters cannot cooperate, they generate independent channel codewords, whereas in the presence of feedback, the transmitters can cooperate through the feedback channel. Indeed, information theoretic random coding schemes that achieve rates outside the capacity region of the classical MAC do enable cooperation by allowing each transmitters to decode each other's message (partially) and collaborate for transmission. We believe our results will open new research directions, since our findings indicate that a fully-learnable framework is not only capable of forming an error-correction mechanism, as in\cite{deepcode, gbaf, baaf, deepcode_act, feedbackcoderobust}, but also able to form a co-operation strategy between the transmitters. 

\section{Preliminaries and Problem definition}

\begin{figure}[t]
\begin{center}
\includegraphics[scale =0.25]{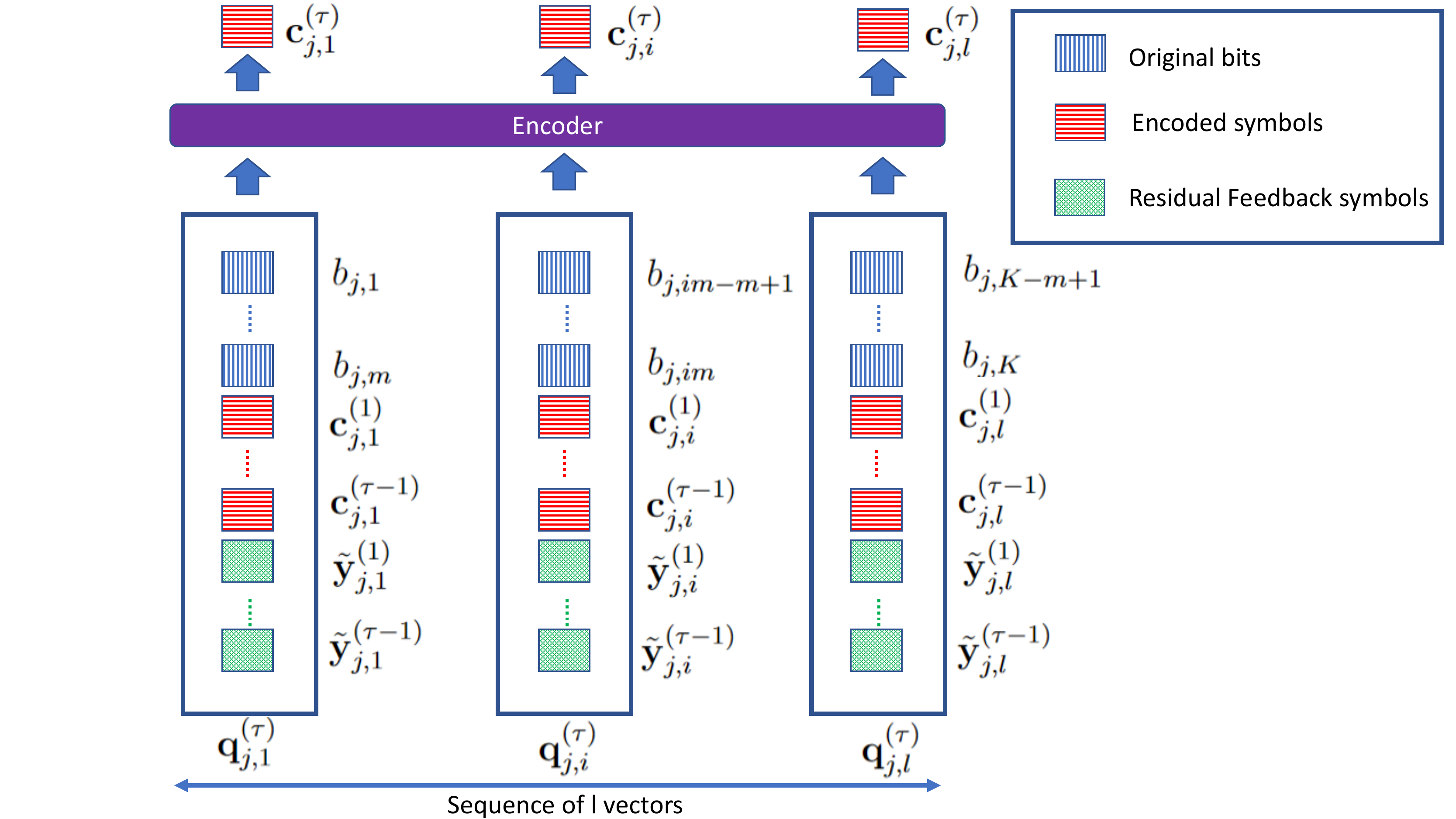}
\caption{Visualisation of sequence-to-sequence encoding with block structure at iteration $\tau$ at the $j$-th user.}
\label{seq2seq}
\end{center}
\end{figure}

\subsection{System Model and Problem Definition}\label{sec:problem_form}
We consider an additive white Gaussian noise (AWGN) MAC with two transmitters. We assume that feedback from the receiver to both transmitters is available in the form of perfect channel output.
We consider a block feedback scenario: Communication is performed over $T$ blocks in the forward direction. After the transmission of each block, the channel output becomes available to the transmitters. During the $\tau$th communication block, $\tau\in[T]$, user/transmitter $j\in\left\{1,2\right\}$ sends $l_{ff}$ symbols over the channel, which is referred to as a block of symbols and denoted  by  $\mathbf{c}^{(\tau)}_{j} = \left[ {c}^{(\tau)}_{j,1}, \ldots, {c}^{(\tau)}_{j,l_{ff}} \right]$. The transmitted symbols are superimposed and exposed to AWGN; that is, the receiver observes the block of noisy superimposed symbols $\mathbf{y}^{(\tau)}$:
\begin{equation}
\mathbf{y}^{(\tau)} = \mathbf{c}^{(\tau)}_{1} +\mathbf{c}^{(\tau)}_{2} + \mathbf{n}^{(\tau)},
\end{equation}
where $\mathbf{n}^{(\tau)}\in\mathbb{R}^{l_{ff}}$  is a vector of independent zero-mean Gaussian random variables with covariance matrix $\sigma^{2}\mathbf{I}_{l_{ff}}$. Due to the perfect channel output assumption, $\mathbf{y}^{(\tau)}$ becomes available to the transmitters as well. 


User $j$ wants to send a $K$-bit message $\mathbf{b}_{j}=[b_{j,1},\ldots,b_{j,K}]\in \left\{0,1\right\}^{K}$ to the receiver in $N = Tl$ channel uses, which corresponds to a \textit{sum-rate} of $R=2K/N$. 

Next, we define the notion of knowledge vector, denoted by $\boldsymbol{q}^{(\tau)}_{j}$, which represents all the available information (history) at the $j$th transmitter at the beginning of block $\tau+1$:

\begin{equation}
    \bm{q}^{(\tau+1)}_{j} \triangleq \left[\bm{b}_{j}, \bm{c}_j^{(1)},\ldots,\bm{c}_j^{(\tau)}, \mathbf{y}^{(\tau)}. \right],
\end{equation}

While we focus on a passive feedback mechanism in the form of perfect channel output feedback, more advanced feedback strategies involving active feedback can be easily incorporated into this framework, following the approach followed in \cite{baaf} for the point-to-point scenario. 

The objective is to design an encoding function $M^{({\tau})}_{j}$, for each transmitter $j$,  to map its knowledge vector to a block of parity symbols at each block, i.e.,
\begin{equation}
M^{({\tau})}_{j}: \boldsymbol{q}^{(\tau)}_{j}   \xrightarrow{} \mathbf{c}^{(\tau)}_{j},
\label{eq:knowledge}
\end{equation}
under an average power constraint imposed for each transmitter:
\begin{align}
\mathbb{E}\left[\frac{1}{N} \sum^{T}_{\tau=1}\langle \mathbf{c}^{(\tau)}_{j} , \mathbf{c}^{(\tau)}_{j} \rangle\right]\leq 1, j\in\left\{1,2\right\},
\label{eq:powernorm}
\end{align}

We also need a decoding function $D$ to be employed at the receiver to estimate $\mathbf{b}_{1}$ and $\mathbf{b}_{2}$ from the knowledge vector $\tilde{\bm{q}}=[\mathbf{y}^{(0)},\ldots,\mathbf{y}^{(T)}]$, i.e.,
\begin{equation}
D: \tilde{\bm{q}} \xrightarrow{} \hat{\mathbf{b}}_{1}, \hat{\mathbf{b}}_{2} \in\left\{0,1\right\}^{K}.
\end{equation}

Hence, the overall objective is to jointly design $M^{({\tau})}_{1},M^{({\tau})}_{2}$ and $D$. In this paper, inspired by \cite{gbaf, baaf}, we propose a DNN-based solution built upon the transformer architecture for the MAC scenario, called the MBAF code. Next, we revisit the fundamental ideas behind the block attention feedback (BAF) code framework introduced in \cite{gbaf, baaf}, and then explain how to adopt it for the MAC scenario. We further introduce a novel {\em sequential decoding} strategy, which decodes the message in multiple iterations, and includes the prediction in the previous iteration as an input to the decoder in order to refine it.

\subsection{Block Attention Feedback Code Structure}
The key idea behind the BAF code design \cite{gbaf,baaf} is to transform the knowledge vector, both at the receiver and the transmitters, into a sequence of vectors, and utilize the self-attention mechanism over these sequences to mimic encoding and decoding operations. Since, from the construction, the self-attention mechanism explores the similarities and differences between the elements of a sequence, while taking into account their positions within the sequence, it is possible to draw an analogy between processing a sequence of vectors and encoding/decoding process of a bit-stream. 

One of the critical aspects highlighted in \cite{gbaf,baaf} is how to form these sequences. The proposed strategy is to divide the $K$ information bits into $l$ blocks each consisting of $m$ bits, i.e., $K = lm$. As a result, the initial bit stream is treated as a sequence of $l$ bit blocks, i.e., $\bm{b}=\left[\bm{b}_{1},\ldots,\bm{b}_{l}\right]$. In parallel with this sequence of bit blocks, the knowledge vector 
$\bm{q}^{(\tau)}$ is also organized into a sequence of $l$ vectors, $\mathcal{Q}^{(\tau)}= \left\{\bm{q}^{(\tau)}_{1}, \ldots, \bm{q}^{(\tau)}_{l} \right\}$, where $\bm{q}^{(\tau)}_{i}$ denotes the knowledge vector corresponding to the $i$th block, and initially we have $\bm{q}_i^{(1)} =  \bm{b}_{i}$. The key design principle of the BAF code is to process the sequence of knowledge vectors $\bm{q}^{(\tau)}$ in a parallel manner, using self-attention, and to generate one encoded symbol for each block. Here, $\bm{c}^{(\tau)}_{i}$ represents the symbol generated corresponding to the $i$th block during iteration $\tau$. Hence, following the reception of the block of feedback symbols, the knowledge vector corresponding to each block is also updated accordingly:
\begin{equation}
    \bm{q}^{(\tau+1)}_{i} = \left[\bm{b}_i, \bm{c}_i^{(1)},\ldots,\bm{c}_i^{(\tau)},  {\bm{y}}^{(1)},\ldots,{\bm{y}}^{(\tau)}\right].
\end{equation}

 The iterative process of updating the knowledge vectors by stacking the new generated symbols and the received feedback symbols, and generating encoded symbols from the sequence of knowledge vectors is continued until a given number of interactions $T$ is reached. Once the the communication is terminated, the decoder decodes the original bit sequence $\bm{b}$ using all its received symbols $\left[ \bm{y}^{(1)}, \ldots, \bm{y}^{(T)} \right]$, again by reshaping each $\bm{y}^{(i)}$ as a sequence of blocks and combining the sequences over blocks. In the next section, we further explain the architectural  details of the BAF code design, and how the sequences are formed with a particular focus on the MAC scenario.

\begin{figure*}[t]
\begin{center}
\includegraphics[scale =0.30]{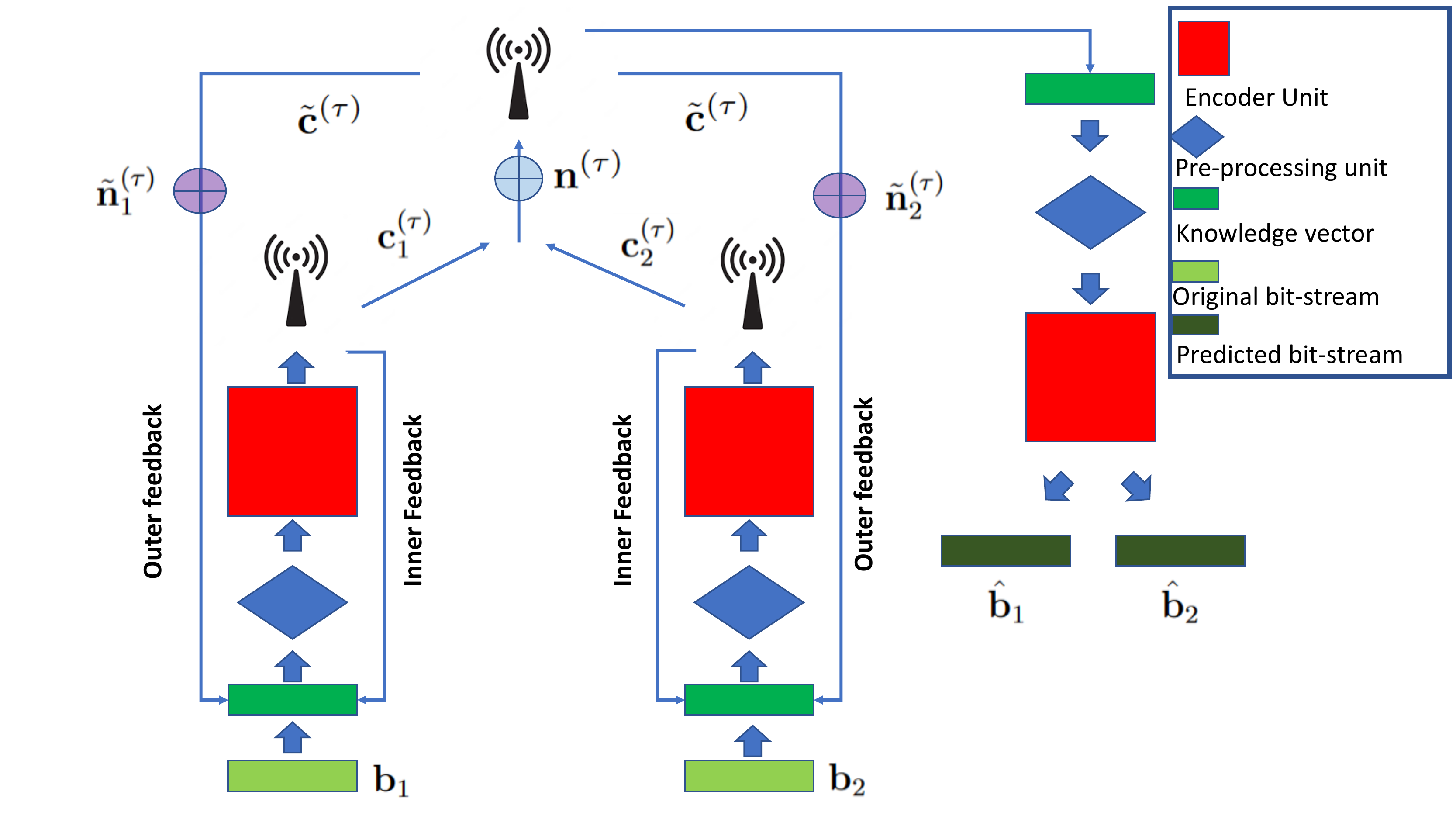}
\caption{Illustration of the overall deep learning assisted communication framework for MAC with feedback scenario.}
\label{sysmodel}
\end{center}
\end{figure*}

\section{Multi-access Block Attention Feedback (MBAF) Codes}

\subsection{Overview of the Design}

As illustrated in Fig. \ref{sysmodel}, the entire design consists of three networks, two \textit{parity networks}, each located at one of the transmitters and a \textit{decoder network} at the receiver. Each network is comprised of similar components: the pre-processing unit, the feature extractor, the sequence-to-sequence encoder and the output mapping, which are denoted as $S(\cdot)$, $H_{extract}, H_{s2s}$ and $H_{map}$, respectively. Among them, the pre-processing unit, $S(\cdot)$, is non-trainable and represents exactly the same operation in all the three networks. The other components, $H_{extract}, H_{s2s}$ and $H_{map}$, however, are parameterized by DNNs and different networks have different realization of these modules. To avoid a possible confusion in the following, we use superscripts `$\mathrm{parity}$' and `$\mathrm{decoder}$' to distinguish the learnable modules employed by the users and the receiver, respectively.
The details for each of the components are described next.

\subsubsection{Pre-processing unit}
$S(\cdot)$ transforms the knowledge vector, $\bm{q}$, into a sequence of vectors for further processing. To be precise, $S(\cdot)$ at the $j$-th user partitions $\bm{q}_j^{(\tau)}$ into $l$ vectors, i.e., $S(\bm{q}_j^{(\tau)}) = \{\bm{q}_{j,1}^{(\tau)}, \ldots, \bm{q}_{j,l}^{(\tau)}\}$ and the operations are the same at the decoder network.

\subsubsection{Feature extractor}
$H_{\mathrm{extract}}$ adopts a multi-layer perceptron (MLP) architecture with a ReLU activation function that transforms $d_{in}$-dimensional input vectors into vectors of dimension $d_{model}$. This component also keeps the output within a specific range to prevent large noise terms. Note that the neural network structure for $H_{\mathrm{extract}}$ is identical in all three networks.

\subsubsection{Sequence-to-sequence encoder}
$H_{s2s}$ takes a sequence of $l$ $d_{model}$-dimensional vectors as input and transforms them into a sequence of $l$ vectors with the same dimension.
We use the transformer architecture as the backbone and denote the output sequence as $\mathcal{V}=\left\{\bm{v}_{1},\ldots,\bm{v}_{l} \right\}$. 

$H_{s2s}$ is comprised of $\bar{N}$ identical {\em transformer encoder layers}, where each layer consists of three primary modules: the feed-forward module, the multi-head attention module, and the layer normalization module as detailed in \cite{gbaf}. In our implementations, we set $d_{model}=32$, and $\bar{N}=2$ for all three networks.

\addtolength{\topmargin}{+0.1in}

\begin{algorithm}[t!]
\caption{MAC parity symbol encoding (MIPSE)}\label{alg:ipse}
\begin{algorithmic}[1]
\For{$\tau=1,\ldots,T$}
\For{\textbf{Transmitter} $j\in \{1,2\}$} Simultaneously
\State{\textbf{Update knowledge vector}:}
\State{$\boldsymbol{q}_j^{(\tau)}=[\bm{b}_{j},\bm{c}_{j}^{(1)},\ldots,\bm{c}_{j}^{(\tau-1)}, \bm{y}^{(1)},\ldots,\bm{y}^{(\tau-1)}]$}
\State{\textbf{Pre-process knowledge vector}:}
\State{$\left\{\bm{q}^{(\tau)}_{j,1},\ldots,\bm{q}^{(\tau)}_{j,l}\right\}=S(\boldsymbol{q}_j^{(\tau)})$}
    \State{\textbf{Feature extraction:}}
    \For{$i\in[l]$}~~$\bm{f}^{(\tau)}_{j,i} = H^{parity}_{extract}(\bm{q}^{(\tau)}_{j,i})$
    \EndFor
    \State{\textbf{Attention-based neural-encoding}:}
    \State{$ \mathcal{V}_j^{(\tau)}= H^{parity}_{s2s}(\mathcal{F}_j^{(\tau)})$}
    \State{\textbf{Symbol mapping}:}
    \For{$i\in[l]$}~~$c^{(\tau)}_{j,i}= H^{parity}_{map}(\bm{v}^{(\tau)}_{j,i})$
    \EndFor
    \State{\textbf{Send block of symbols} $\mathbf{c}_j^{(\tau)} = \left[c^{(\tau)}_{j,1}, \ldots, c^{(\tau)}_{j,l}\right]$}
\EndFor
\State{\textbf{Receiver}: broadcast ${\bm{y}}^{(\tau)}$ while $\tau < T$.}
\EndFor
\end{algorithmic}
\end{algorithm}

\subsubsection{Output mapping}
$H_{\mathrm{map}}$ maps the latent representation $\mathcal{V}=\left\{\bm{v}_{1},\ldots,\bm{v}_{l} \right\}$  to the desired output $\bm{u}_{i} = H_{\mathrm{map}}(\bm{v}_{i})$, for all $i\in[l]$. For all the three networks, $H_{map}$ is parameterized using an MLP layer. Specifically, the output parity symbols at the $j$-th user are ${u}_{j,i}=c_{j,i} = H_{map}^{parity}(\bm{v}_{j,i}), i\in [l]$ with $d_{out}=1$ while the decoder network at the receiver maps the latent representation $\bm{v}_i$ to two probability vectors as $\bm{u}_{i} = \{\bm{w}_{1,i}, \bm{w}_{2,i}\} = H^{decoder}_{map}(\bm{v}_i)$ with $d_{out}=2^{m}$, where $\bm{w}_{j,i}$ represents the probability vector corresponding to the $i$-th bit block of the $j$-th user. Note that the MLP in $H^{decoder}_{map}$ is followed by an additional softmax layer. Finally, the decoded bits are estimated from the probability vectors for each user.

\begin{algorithm}[t!]
\caption{MAC joint successive parity symbol decoding (MJSPSD)}\label{alg:jpsd}
\begin{algorithmic}[1]
\State{\textbf{Initialize knowledge vector}:}
\State{$\hat{\boldsymbol{q}}^{(0)} =[\bm{y}^{(1)},\ldots,\bm{y}^{(T)}]$}
\State{\textbf{Pre-process knowledge vector for decoder network}:}
\State{$\left\{\hat{\bm{q}}_{1}^{(0)},\ldots,\hat{\bm{q}}_{l}^{(0)}\right\}=S(\hat{\boldsymbol{q}}^{(0)})$,~~ $\hat{\bm{q}}_{i}^{(0)}=[{\bm{y}}^{(1)}_{i},\ldots,{\bm{y}}^{(T)}_{i}])$}
\State{\textbf{Feature extraction:}}
\For{$i\in[l]$}~~${\bm{f}}_{i}^{(0)} = H^{decoder}_{extract}(\hat{\bm{q}}_{i}^{(0)})$
\EndFor
\State{\textbf{Attention-based decoding}: $ {\mathcal{V}}^{(0)}= H^{decoder}_{s2s}({\mathcal{F}}^{(0)})$}
\State{\textbf{Mapping}:}
\For{$i\in[l]$}
$\{\bm{w}_{1,i}^{(0)}, \bm{w}_{2,i}^{(0)}\}= H^{decoder}_{map}({\bm{v}}_{i}^{(0)})$
\EndFor
\For{$n\in[N_{iter}]$}
\State{\textbf{Convert class-wise prediction to bit-wise belief}:}
\For{$i\in[l]$}
$\{\bar{\bm{b}}_{1,i}^{(n)}, \bar{\bm{b}}_{2,i}^{(n)}\}= \{\bm{A}\bm{w}_{1,i}^{(n)}, \bm{A}\bm{w}_{2,i}^{(n)}\}$
\EndFor
\State{\textbf{Enhance the knowledge vectors }:}
\State{$\hat{\bm{q}}_{i}^{(n)}=[{\bm{y}}^{(1)}_{i},\ldots,{\bm{y}}^{(T)}_{i},\bar{\bm{b}}_{1,i}^{(n)}, \bar{\bm{b}}_{2,i}^{(n)}])$}
\State{\textbf{Sequence-to-sequence Decoding:} Repeat lines 5-9}
\EndFor
\State{\textbf{Block-wise classification}:} 
\For{$i\in[l], j\in\{1,2\}$} $I_{j,i}= arg\,max(\bm{w}_{j,i}^{(N_{iter})})$
\EndFor
\State{\textbf{Blockwise index to bitstream conversion}:}
\For{$i\in[l], j\in\{1,2\}$}  $\hat{\bm{b}}_{j,i}=f_{d2b}(I_{j,i})$
\EndFor
\State{\textbf{Retrieve the bitstream of each user}:}
\For{$j\in\{1,2\}$}  $\hat{\bm{b}}_{j}=\left[\hat{\bm{b}}_{j,1}, \ldots, \hat{\bm{b}}_{j,l} \right]$
\EndFor
\end{algorithmic}
\end{algorithm}

\subsection{Algorithmic Flow and Training Procedure}
In this subsection, we explain the end-to-end encoding and decoding processes of the proposed MBAF code, which involves two algorithmic flows: {\em MAC iterative parity symbol encoding (MIPSE)} and {\em MAC joint successive parity symbol decoding (MJSPSD)}.

MIPSE, summarized in Algorithm \ref{alg:ipse}, illustrates the encoding process at the users. 
As shown in Fig. \ref{seq2seq}, the exact form of the knowledge vector in \eqref{eq:knowledge} is given as $\boldsymbol{q}_j^{(\tau)}=[\bm{b}_{j},\bm{c}_{j}^{(1)},\ldots,\bm{c}_{j}^{(\tau-1)}, \tilde{\bm{y}}_j^{(1)},\ldots,\tilde{\bm{y}}_j^{(\tau-1)}]$, where $\bm{c}_{j}^{(i)}$ denotes the already transmitted symbol and $\tilde{\bm{y}}^{(i)}_j = \bm{y}^{(i)} - {\bm{c}}^{(i)}_j$ is referred to as the residual feedback symbols for $i\in [1, \tau-1]$. The algorithm iteratively generates transmit symbols $\bm{c}_{j}^{(\tau)}$ by exploring the feedback from the receiver, and terminates when $T$ iterations are reached. 

After the transmission is completed at the $T$-th communication block, the receiver employs the MJSPSD  algorithm as summarized in Algorithm \ref{alg:jpsd}. The algorithm takes the received signal as input and outputs the predicted bit sequences $\hat{\bm{b}}_1, \hat{\bm{b}}_2$, for each of the two users. To improve the error correction ability of the decoder network, we use successive decoding with $N_{iter}>0$ iterations. The decoding process starts by generating the latent probability vectors $\{\bm{w}_{1,i}^{(0)}, \bm{w}_{2,i}^{(0)}\}$ following the same procedure as MIPSE, except the outputs $\bm{w}_{j,i}^{(0)}$ represent the probability vectors with dimension $2^m$.  We transform these vectors to belief vectors $\bar{\bm{b}}_{j,i}^{(0)} \in [0, 1]^m$ by multiplying them with a binary matrix $\bm{A}$ with dimension $m\times 2^m$, which can be expressed as:
\begin{align}
    \bar{\bm{b}}_{j,i}^{(0)} &= \bm{A}\bm{w}_{j,i}^{(0)},
    \label{eq:belief_vector}
\end{align}
where $\bm{A}$ for $m=3$ is given as: 
\begin{align}
\begin{pmatrix} 
0 & 0 & 0 & 0 & 1 & 1  & 1 & 1\\  
0 & 0 & 1 & 1  & 0 & 0 & 1 & 1 \\ 
0 & 1 & 0  & 1 & 0 & 1 & 0 & 1 \\ 
\end{pmatrix}.
\end{align} Note that the $k$-th entry of $\bar{\bm{b}}_{j,i}^{(0)}$ indicates the probability of $\bm{b}_{j,i}[k] = 1$. For the $n$-th iteration, we update the knowledge vector by concatenating the received signal $\bm{y}$ with the belief vectors $\bar{\bm{b}}_{j,i}^{(n)}$, and following the same procedure of MIPSE to generate refined probability vectors $\{\bm{w}_{1,i}^{(n)}, \bm{w}_{2,i}^{(n)}\}$. Finally, after $N_{iter}$ number of iterations, we obtain the index of the largest value in $\bm{w}_{j,i}^{(N_{iter})}$, denoted by $I_{j,i}$, and transform it into the bit sequence $\bm{\hat{b}}_{j,i}$, using a decimal-to-binary mapping $f_{d2b}(\cdot)$. By collecting $\bm{\hat{b}}_{j,i}, i\in [l]$ for each user and concatenating them, the final estimated bit sequences $\{\hat{\bm{b}}_1, \hat{\bm{b}}_2\}$ are obtained. 

Next, we will introduce the training methodology for the proposed MBAF code. The initial step is to randomly generate two bit sequences $\bm{b}_j\in\left\{0,1\right\}^{K}$ for each user, then divide each bit sequence into $l$ blocks, each consisting of $m$ bits. The corresponding label is assigned to each block and the training data is organized as $\mathcal{B}_j= \left\{(\bm{b}_{j,1},y_{j,1}), \ldots, (\bm{b}_{j,l},y_{j,l})\right\}, j\in\{1,2\}$. Next, the generated data is fed to the autoencoder, which produces a sequence of $l$ probability vectors for each user: $\left\{\bm{w}_{j,1}^{(N_{iter})}, \ldots,\bm{w}_{j,l}^{(N_{iter})}\right\}$. The cross-entropy loss is adopted and the total loss is averaged over the users: 
\begin{equation}
L_{j}(\bm{W}_{j},Y_{j}) = \sum^{l}_{i=1}\sum^{2^{m}-1}_{c=0}-\log{{(\bm{W}_{j}{[i,c]})}}\cdot\mathbbm{1}_{y_{j,i}\neq c}
\end{equation}
where $Y_{j}=\left\{y_{j,1},\ldots,y_{j,l}\right\}$ is the list of labels and $\bm{W}_{j}$ is the matrix form of the sequence $\{\bm{w}_{j,1}^{(N_{iter})},\ldots,\bm{w}_{j,l}^{(N_{iter})}\}$ for the $j$-th user.

\begin{table*}[t]
\centering
\caption{Average BLER performance of MBAF code for different \textit{sum-rate} $R$ values.}
\label{bler}
\begin{tabular}{@{}lllllll@{}}
\toprule
\textbf{SNR/Rate} &  6/12  &  6/11 &  6/10 & 6/9 & 6/8 & 6/7\\ \midrule
-1 dB & $2.6\times10^{-8}$ & $5.9\times10^{-8}$ & $3.1\times10^{-5} $ & $3.5\times10^{-2}$ & -&-\\
-0.5 dB & - & $1.7\times10^{-8}$  & $8.8\times10^{-8}$ & $4.8\times10^{-5}$ & $8.2\times10^{-2}$ & - \\
0 dB & - & - & $6\times10^{-10}$ & $6.2\times10^{-7}$ & $2.9\times10^{-4}$ & - \\
0.5 dB & - &  - & - & $1.7\times10^{-8}$  & $7.5 \times10^{-6} $ & $5\times10^{-2} $  \\
1 dB & - & - & - & $2.9\times10^{-9}$ & $1.7\times10^{-7}$  & $6.7\times10^{-3}$ \\ 
1.5 dB & - & - & - & $7.5\times10^{-9}$  & $4.48\times 10^{-8}$ & $1.4\times 10^{-4}$ \\ 
\end{tabular}
\end{table*}

\section{Numerical Results}\label{s:experiments}
To evaluate the performance of the proposed MBAF code, we consider each user transmitting a bit stream of length $K=51$ bits and set the number of blocks $l=17$ with block size $m = 3$. In all the experiments, we consider the number of iterations $N_{iter} = 2$ for a better BLER performance.\\
\indent In the training phase, we use a large batch size of $B=8192$ and employ the AdamW optimizer with an initial learning rate of $0.001$ and a weight decay parameter $0.01$. The gradient values are clipped with a threshold value of $0.5$ for stable training. The model is trained using $180K$ batches. Finally, we follow curriculum learning to train the models for low target signal-to-noise ratio (SNR) values, where the SNR of the forward channel ($SNR_{ff} \triangleq 1/\sigma^2$) decreases linearly from $3$ dB to the target value in the first $30K$ batches. 

We evaluate the BLER performance of the proposed MBAF code for $SNR_{ff}\in \{-1, -0.5, 0, 0.5, 1, 1.5\}$~dB, and  the number of interactions/blocks $T=\{7,\ldots,12\}$, which corresponds to \textit{sum-rate} values $R = \{6/12,\ldots, 6/7\}$. The overall BLER is defined by the average BLER of the two users, $\text{BLER} = \frac{1}{2}\mathbb{E}(\mathbbm{1}_{\hat{\bm{b}}_1 \neq {\bm{b}}_1} + \mathbbm{1}_{\hat{\bm{b}}_2 \neq {\bm{b}}_2})$ and the results for different settings are illustrated in Table \ref{bler}.

As a benchmark, we also consider a time-division multiple access (TDMA) scheme, where each user, with perfect feedback, transmits its message only in its assigned time slot. We consider a simplified case where $T$ is an even number and each user occupies $T/2$ number of interactions; that is, they equally share the available bandwidth. We use the GBAF code proposed in \cite{gbaf} to generate the BLER of the TDMA scheme, denoted by GBAF-TD.  Note that for a fair comparison, in GBAF-TD, $SNR_{ff}$ is increased by $3$ dB to ensure that the total energy consumption is the same. In this simulation, we set $SNR_{ff} = \{-1, 0, 1, 2\}$ dB and $T = 8, K = 51$.  As shown in Table \ref{tab:TD_vs_MBAF}, the BLER of the proposed MBAF significantly outperforms the GBAF-TD in all $SNR_{ff}$ values.

\begin{figure}[t]
\begin{center}
\includegraphics[scale=0.4]{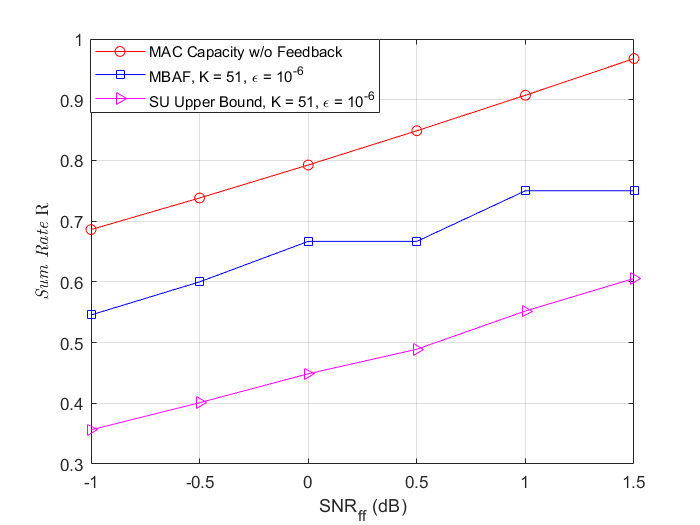}
\caption{Illustration of the \textit{sum rate}, $R$, of the MBAF code with a BLER $\epsilon = 10^{-6}$ compared to the theoretical \textit{sum rate} of the multiple access channel without feedback both in finite block length and infinite block length (capacity).}
\label{rateSNR}
\end{center}
\end{figure}

As another benchmark, we consider the upper bound on the achievable performance for a point-to-point channel without feedback in the finite block length regime \cite{finite_length_capacity}, which can be obtained by treating the two users as a `super user' that uses twice the power over $T$ interactions. In this simulation, we consider $K = 51, \epsilon = 10^{-6}$ for both the proposed MBAF code and the point-to-point benchmark, where $\epsilon$ is the BLER threshold which is an indicator of reliable communication. The achievable \textit{sum-rate} $R$ of the two schemes with respect to different $SNR_{ff}\in [-1, 1]$ dB are shown in Fig.  \ref{rateSNR}. Note that we also provide the MAC sum capacity without feedback, which is given by $\frac{1}{2}\log_2(1 + \frac{2P}{\sigma^2})$. As shown in Fig. \ref{rateSNR}, the proposed MBAF scheme outperforms the achievable rate upper bound of the point-to-point baseline without feedback in the finite block length, which confirms the gain brought by exploiting feedback from the receiver. The achieved rate is below the MAC sum-capacity, but this is expected as we are considering an extrmeely short blocklength regime.

\begin{table}[t]
\centering
\caption{BLER comparison of the GBAF-TD  with the proposed MBAF with $K = 51, R = 6/8$.}
\label{tab:TD_vs_MBAF}
\begin{tabular}{@{}lllll@{}}
\toprule
\textbf{SNR} &  $-1$ dB  &  $0$ dB &  $1$ dB\\ \midrule
GBAF-TD & $8.4\times10^{-1}$ & $3.6 \times 10^{-1}$ & $1.8\times10^{-3}$\\
MBAF & $4.0 \times 10^{-1}$ & $2.9 \times 10^{-4}$ & $1.7\times10^{-7}$ \\
\bottomrule
\end{tabular}
\end{table}

\section{Conclusion}\label{s:Conclusion}
In this paper, we proposed the MBAF code for reliable communication over a MAC with perfect channel output feedback. The MBAF code is designed in a purely data-driven manner. The parity symbols are generated sequentially at each user, using a block-based communication framework, and employing transformer-based neural network models. Our results confirm that the users learn not only to mitigate the adverse effects of the channel noise, but also to cooperate among each other by utilizing the feedback signal. This is a first step towards utilizing the transformer architecture to design novel channel codes in highligh challenging multi-user scenarios.

\bibliographystyle{IEEEtran}
\bibliography{IEEEabrv,ref.bib}

\end{document}